# STREAMING-CAPABLE HIGH-PERFORMANCE ARCHITECTURE OF LEARNED IMAGE COMPRESSION CODECS

Fangzheng Lin[1]
lin_toto@toki.waseda.jp

Heming Sun[2,3]
hemingsun@aoni.waseda.jp

Jiro Katto[1,2]
katto@waseda.jp

[1] School of Fundamental Science and Engineering, Waseda University, Tokyo, Japan
[2] Waseda Research Institute for Science and Engineering, Waseda University, Tokyo, Japan
[3] JST, PRESTO, 4-1-8 Honcho, Kawaguchi, Saitama, Japan

**ABSTRACT**

**Learned image compression allows achieving state-of-the-art accuracy and compression ratios, but their relatively slow runtime performance limits their usage. While previous attempts on optimizing learned image codecs focused more on the neural model and entropy coding, we present an alternative method to improving the runtime performance of various learned image compression models. We introduce multi-threaded pipelining and an optimized memory model to enable GPU and CPU workloads' asynchronous execution, fully taking advantage of computational resources. Our architecture alone already produces excellent performance without any change to the neural model itself. We also demonstrate that combining our architecture with previous tweaks to the neural models can further improve runtime performance. We show that our implementations excel in throughput and latency compared to the baseline and demonstrate the performance of our implementations by creating a real-time video streaming encoder-decoder sample application, with the encoder running on an embedded device.**

## I. INTRODUCTION

Learned image compression methods have been gaining interest nowadays for the superior image quality to its compression ratio. In recent years, multiple state-of-the-art learned image compression codecs were introduced, for example, [1], [2], [3], [4], [5], [15] and [20]. While these models offer exceptional accuracy compared to traditional methods, the tremendous amount of hardware resource that learned image compression require, which limits their throughput and latency, makes their usage not often seen, especially for streaming purposes and on embedded devices.

We present an approach for implementing learned codecs, demonstrating with a factorized-prior [1] and a scale-hyperprior [2] codec. We show that: (1) by employing practical implementation techniques, namely multithreaded pipelining, zero-copy memory, and memory pooling, we can speed up the codecs without modifying either the learned models or the entropy coders; (2) by combining our architecture with other works on optimizations of the learned models (e.g., [6]), the codecs can reach excellent performance; (3) a demonstration of a learned image compression based video monitoring system, with the embedded platform Jetson Xavier NX [10] as an encoder, serving a real-time stream of 1280x720 at 30 FPS video feed to a server, to show the performance of our implementation.

## II. RELATED WORK

Previous attempts to improve the runtime performance of learned image codecs focused on either the neural model architecture or the entropy coding process. For example, [7] featured a codec architecture designed with real-time operation in consideration. [14] designed a block-based encoder architecture that splits the input image into blocks for more flexibility in parallel context prediction. [6] demonstrated using fast activation layers and model pruning to reduce computational complexity on existing models.

On the other hand, works like [16] and [19] showed different approaches in implementing high-performance entropy coders. [17] and [18] demonstrated different methods enabling parallel entropy parameter calculations in context models, significantly boosting the serial masked CNN performance bottleneck. These works showed performance increases of the entropy coding process of different image codecs.

However, there have been no real-time implementations of learned image compression demonstrated in video rate (i.e., 30-60 fps), partly because the often ignored third bottleneck of learned image compression systems: the integration and coordination between the neural model and entropy coding as a system. The neural model and entropy coding require very different computing resources (often GPU and CPU, respectively). In a direct implementation such as [8], GPU and CPU workload are serially executed. This is not ideal as when GPU performs computation, the CPU waits doing nothing, and vice versa. It results in very poor resource usage and makes the optimization on either half soon reach diminishing returns.

Our architecture resolves this bottleneck by enhancing coordination between GPU and CPU. Unlike previous works, our method works on most existing learned image codecs without either new neural model or entropy coder architectures. Our approach also comes with the capability to be further optimized when combined with the above model optimizations such as [7], [14], and [6], or high-performance entropy coders such as [16] and [19].

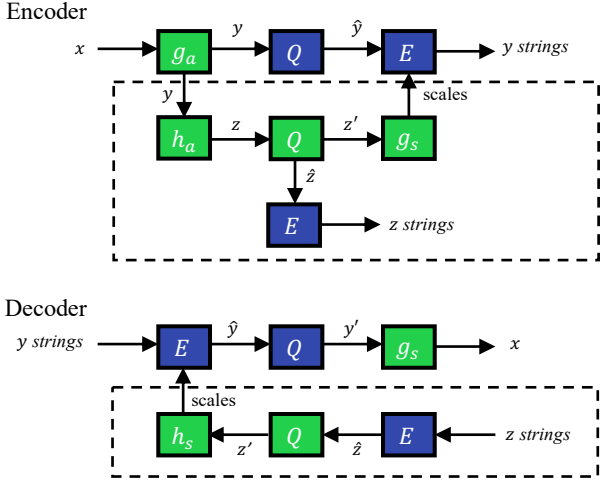

Fig. 1. Operation diagram of the encoder and decoder of the two codecs. Dotted square areas represent scale-hyperprior's extension to factorized-prior. Green squares represent GPU workloads; Blue squares represent CPU workloads. Q represents (de-)quantization; E represents entropy (de-)coding.

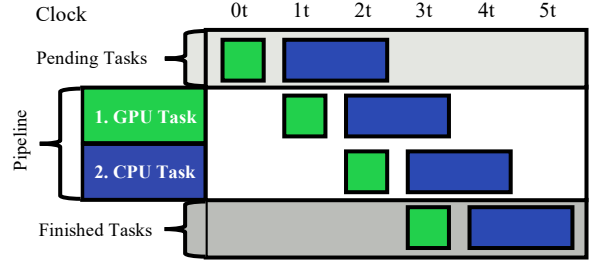

Fig. 2. Timing diagram of a 2-stage encoder pipeline

We demonstrate our architecture with the factorized-prior [1] and scale-hyperprior [2] codec for simplicity. We run these neural models on GPUs as they are nowadays readily available even on some embedded devices and the corresponding entropy coders on CPU. As mentioned, our method could also be extended to support other codecs or other types of neural workloads that run on FPGA/ASIC.

## III. ARCHITECTURE

We performed mainly the following optimizations in our architecture.

### A. Asynchronous workloads for factorized-prior

As Figure 1 shows, the factorized-prior [1] model encodes an image in three steps: (1) non-linear transformation of the image $x$ into a transform domain, outputting $y$; (2) quantization of $y$ with a channel-wise mean, producing $\hat{y}$; and (3) entropy coding of $\hat{y}$. The transformation is a computationally heavy but highly scalable process and performs best if executed on the GPU. The quantization process has only a negligible performance impact. The entropy coding process is highly CPU-intensive.

We introduce a multi-threaded pipeline framework into the architecture. As previously explained, the factorized-prior learned image codec can be split into a GPU-intensive and a CPU-intensive workload. Each frame of the input image (or strings for decoder) is represented by a state-machine pattern task, which must be processed by the above workloads sequentially. For each workload, a FIFO task queue implementing a multi-threaded producer-consumer pattern manages the execution of the tasks. A raw frame enters the pipeline from the rear of the first workload's task queue. One or more threads would fetch tasks from the queue and execute the workload: for a GPU-intensive workload, a dedicated control thread fetches the tasks and communicate with the GPU; for a CPU-intensive workload, multiple worker threads perform the computations. When a task finishes computation, its result is put directly into the queue of the next task. Therefore, by coordinating the execution of a learned image codec, where each frame travels through the pipeline, it is processed by the workloads sequentially; in contrast, the processing of different frames happens parallelly, increasing overall throughput and resource usage.

For factorized-prior encoder and decoder [1], the GPU-intensive workload is the transformation $g_a$ / inverse transformation $g_s$ model, and the CPU workload performs (de-)quantization and entropy (de-)coding. We implement this encoder and decoder with a 2-stage pipeline. Figure 2 shows the timing diagram of this encoder.

### B. Asynchronous workloads for scale-hyperprior

As Figure 1 shows, the scale-hyperprior model extends the factorized-prior model with a new random variable $z$ carrying spatial dependency information inferred from $y$, through a model $h_a$. The extra data $z$ also needs to be quantized and entropy coded while the entropy coder of $\hat{y}$ requires scales information deduced from $z$ with another model $h_s$ [2].

The scale-hyperprior encoder, while structurally more complicated, is also implemented in a 2-stage pipeline, shown in Figure 2: we combine the models $g_a$, $h_a$, the quantization of $z$ and model $h_s$ into a single runtime, the GPU-intensive workload; and we launch quantization of $y$ and entropy coding of outputs $y$ and $z$ on CPU.

As Figure 1 shows, the scale-hyperprior decoder is more complicated. First, the entropy decoding of $z$ strings is executed on the CPU. The result $z$ is then passed to the neural model $h_s$ to generate the scales on GPU. The scales are needed for the entropy decoding and dequantization of $y$ strings, executed on the CPU. Finally, the decoding result $y$ is used by the model $g_s$ to infer the result image $x$ on GPU. Therefore, instead of a 2-stage pipeline, we implement this decoder with a 4-stage pipeline: CPU1-GPU1-CPU2-GPU2.

### C. Use of TensorRT as inference runtime

Converting the learned model into NVIDIA TensorRT [11] format and using it as inference runtime has various benefits. For example, it compiles the model into native CUDA kernels and eliminates the overhead brought by machine learning frameworks. We found that the use of TensorRT gives an improved performance on the neural models over the sample implemented with PyTorch [8].

### D. Zero-copy memory and memory pooling

TensorRT also enables access to low-level CUDA-specific features, the most important being CUDA zero-copy memory. As most embedded devices use a unified memory for CPU and GPU, the zero-copy memory model eliminates most data copy overhead. This is, however, non-ideal on desktop architectures since GPU and CPU have their dedicated memory. Thus, we offer a command-line option in our implementation to quickly toggle this option on or off.

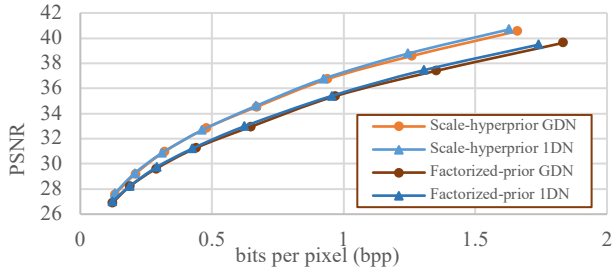 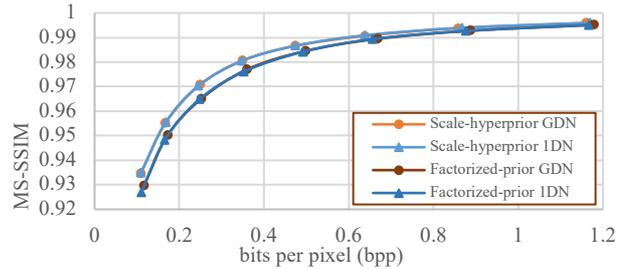

Fig. 3. Left: Rate-distortion curves of the PSNR optimized models. Right: Rate-distortion curves of the MS-SSIM optimized models.

TABLE 1. $\lambda$ VALUES OF 1DN MODELS

| Quality | Architecture | | | |
|---|---|---|---|---|
| | Factor PSNR | Factor SSIM | Hyper PSNR | Hyper SSIM |
| 1 | 0.0016 | 2.30 | 0.0017 | 2.40 |
| 2 | 0.003 | 4.35 | 0.0032 | 4.58 |
| 3 | 0.0055 | 8.29 | 0.006 | 8.30 |
| 4 | 0.01 | 14.98 | 0.0115 | 15.80 |
| 5 | 0.019 | 28.58 | 0.023 | 30.14 |
| 6 | 0.04 | 54.00 | 0.0445 | 57.47 |
| 7 | 0.075 | 101.52 | 0.086 | 109.60 |
| 8 | 0.14 | 190.00 | 0.165 | 210.00 |

TABLE 2. THROUGHPUT AND LATENCY COMPARISON OF THREE IMPLEMENTATIONS, ON 768X512 IMAGES AND JETSON XAVIER NX

| | | Factor Enc | Factor Dec | Hyper Enc | Hyper Dec |
|---|---|---|---|---|---|
| **FPS** | A | 20 | 17 | 13 | 14 |
| | B | 75 | 35 | 68 | 34 |
| | C | 88 | 35 | 83 | 34 |
| **Latency / ms** | A | 51 | 56 | 74 | 67 |
| | B | 22 | 37 | 33 | 44 |
| | C | 20 | 36 | 31 | 44 |

We also find in our profiling that CUDA memory allocation and deallocation can sometimes become a significant overhead. Specifically, we noticed that the `cudaFree` call, which releases memory used by the GPU, blocks until all current CUDA kernels finish executing, causing degradation in throughput and latency. This has an especially devastating effect on the scale-hyperprior 4-stage decoder pipeline; if the GPU1 workload (relatively fast) finishes processing and wants to release some memory, it must wait until the completion of the currently executing GPU2 workloads (relatively slow), causing latency not only to climb but also jitter randomly. Therefore, we carefully control dynamic memory allocation/deallocation and pool the allocated memory for later re-use whenever possible.

## IV. IMPLEMENTATIONS

The factorized-prior and scale-hyperprior models from CompressAI [8] have 8 different quality levels, each corresponding to a λ parameter which controls the loss function into favoring various bits per pixel settings, and optimizing for either PSNR or MS-SSIM. We found that if the model structure (i.e., the channel numbers for convolution and entropy bottleneck) is the same, the quality levels, optimization targets and/or different input images have no impact on throughput and only a minor impact on entropy coder latency. Our performance evaluation focuses on the model with quality parameter 2 optimized for PSNR, with 128 channels for convolution and 192 channels for entropy bottleneck.

Then, we evaluate the following three implementations:

### A. Baseline

We first evaluate the factorized-prior and scale-hyperprior codecs using the models trained by CompressAI and their sample implementation [8]. To ensure a fair comparison, we modify it to infer in FP16.

### B. Pipelined, GDN

We then evaluate the above codecs, using the same neural models converted into TensorRT FP16 format, and with the codec re-implemented using our architecture. We do not change the neural model or perform any other optimizations other than the above specified. For entropy coder, we do not perform optimizations as well and simply integrate the CompressAI entropy coder [8] into our system.

### C. Pipelined, 1DN

We base this implementation on B (Pipelined, GDN) but introduce an optimized model. We discovered a large portion of the performance overhead lies within the GDN activation function [9]. Using the 1DN activation layer [6] provides a considerable speed-up without accuracy loss.

We trained in total 32 variations of tweaked models for both factorized-prior and scale-hyperprior. The λ values are shown in Table 1 (We abbreviate factorized-prior as "Factor" and scale-hyperprior as "Hyper"). For quality 1 to 5, we use 128 channels for convolution and 192 for entropy bottleneck; for 6 to 8 we use 192 and 320 channels respectively; these channel numbers are the same as the CompressAI models [8]. We trained the factorized-prior models on a subset of the Open Image Dataset [12] of about 100,000 randomly selected images, and the scale-hyperprior models on a subset of 300,000 images as we found the models overfit on the smaller subset. We evaluate the accuracy of our models against the baseline on the Kodak Image Dataset [13].

As Figure 3 shows, replacing the GDN activation layer with 1DN does not incur any performance loss for both PSNR and MS-SSIM optimized models. Although only by a small margin, our models optimized for PSNR outperform the baseline; however, this is likely due to the change in the dataset used.

## V. EXPERIMENTS

### A. Performance evaluation on embedded platform

We evaluate the runtime performance of the implementations on Jetson Xavier NX [10], an embedded platform provided by NVIDIA. We give a 768x512 and a 1280x720 sample image as input in an infinite loop and then measure the throughput and latency at the output side. We launch 3 threads for the entropy coder workload.

TABLE 3. THROUGHPUT AND LATENCY COMPARISON OF THREE IMPLEMENTATIONS, ON 1280X720 IMAGES AND JETSON XAVIER NX

|  |  | Factor Enc | Factor Dec | Hyper Enc | Hyper Dec |
|---|---|---|---|---|---|
| FPS | A | 9 | 10 | 7 | 8 |
|  | B | 30 | 16 | 28 | 15 |
|  | C | 37 | 16 | 33 | 15 |
| Latency / ms | A | 110 | 101 | 140 | 127 |
|  | B | 53 | 84 | 85 | 104 |
|  | C | 49 | 83 | 79 | 106 |

TABLE 4. THROUGHPUT AND LATENCY COMPARISON OF THREE IMPLEMENTATIONS, ON 1280X720 IMAGES AND DELL T630

|  |  | Factor Enc | Factor Dec | Hyper Enc | Hyper Dec |
|---|---|---|---|---|---|
| FPS | A | 15 | 19 | 11 | 16 |
|  | B | 316 | 166 | 279 | 157 |
|  | C | 328 | 166 | 281 | 161 |
| Latency / ms | A | 67 | 52 | 86 | 61 |
|  | B | 25 | 22 | 36 | 33 |
|  | C | 26 | 22 | 36 | 37 |

As Table 2 shows (We abbreviate factorized-prior as "Factor" and scale-hyperprior as "Hyper", encoder as "Enc" and decoder as "Dec"), for 768x512 images, the pipelining method alone brings 3.75x and 5.2x throughput boost to factorized-prior and scale-hyperprior encoders, respectively. The additional 1DN activation layer optimization brings another 17% and 23% throughput boost to the two encoders. The pipelining also brings an almost 2x boost to factorized-prior decoder throughput and 2.4x to the scale-hyperprior decoder. The optimizations also reduce the latency from the baseline to about 0.45x for both encoders and 0.65x for both decoders; however, the 1DN activation provides only a marginal benefit for encoder latency and almost no benefit to the decoders.

The same improvement pattern can also be seen on 1280x720 images shown in Table 3. It is clearly demonstrated that with our architecture and the 1DN optimization, a Jetson Xavier NX is more than capable of encoding a live 1280x720 at 30 FPS video feed.

*B. Performance evaluation on desktop platform*

We evaluate the runtime performance of the implementations on a Dell T630 server, with dual Xeon E5-2680 v4 CPU and a NVIDIA GeForce RTX 2080 Ti GPU. We experiment with 1280x720 images as input for the desktop platform. We launch 10 threads for the entropy coder workload.

As Table 4 shows, our architecture alone boosts encoder throughput by 21x and 25x, and decoder throughput by 8.7x and 9.8x, for factorized-prior and scale-hyperprior, respectively. This further demonstrates the importance of good coordination between different hardware. However, the 1DN optimization does not show a significant boost in this experiment. Since the RTX 2080 Ti is a very high-performance GPU, the coordination strategies between the workloads themselves, such as mutexes and queue operations, unfortunately also become part of the overhead in such high throughput situations. The effect of our architecture on encoder/decoder latency is similar to the previous experiment.

VI. STREAMING DEMONSTRATION

Based on the high-performance architecture of the learned image compression models, we create a sample video streaming application to demonstrate the performance further.

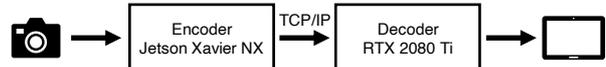

Fig. 4. Architecture of the streaming system

We simulate a video monitoring system setup composed of capturing devices consisting of a camera and an embedded processing board (we use only one capturing device in this experiment.) The devices transfer streams of video encoded by learned image compression to a central server over TCP/IP networking.

The capturing side is a Jetson Xavier NX board with a CSI camera attached, encoding a 1280x720 stream at 30 FPS. The stream is encoded by the factorized-prior model with 1DN activation. Due to the nature of the model, we encode every frame as keyframes. The receiving side is the above Dell T630 server, with RTX 2080 Ti. Due to the decoder of this learned image codec requires more computational power than the encoder, we run it on more powerful hardware; this however exactly represents the use case of a video monitoring system.

Our implementation successfully transmits the camera video stream with no frame drops or noticeable jittering[1].

VII. CONCLUSIONS

Our proposed implementations of learned image compression successfully accelerate the factorized-prior and scale-hyperprior image codecs, pushing their runtime performance to streaming-capable levels even on embedded devices. This is a giant leap on the path to enabling the use of learned image compression in more performance-critical situations.

Even with our architecture, learned image encoding and decoding is still a highly GPU-bound algorithm on most platforms. Therefore, as future work, by combining our method with other research to optimize the neural model, even higher performance could be achieved; on the other hand, the introduction of higher performance entropy coders such as [16] and [19] could potentially further reduce latency on desktop platforms. Especially for context-based models, combining our architecture with works such as [17] and [18] enables more parallel execution and higher throughput to be achieved.

ACKNOWLEDGMENT

This work was supported in part by NICT, Grant Number 03801, Japan; in part by JST, PRESTO Grant Number JPMJPR19M5, Japan; in part by Japan Society for the Promotion of Science (JSPS), under Grant 21K17770; and in part by Kenjiro Takayanagi Foundation.

---

[1] Demo video available at: https://youtu.be/1HJpYPb8fOo.